\documentclass[twocolumn,showpacs,preprintnumbers,prl,amsmath,amssymb]{revtex4}
\usepackage{graphicx}% Include figure files
\usepackage{dcolumn}% Align table columns on decimal point
\usepackage{bm}% bold math

\begin{document}

%\preprint{APS/123-QED}

\title{Dynamic polarization of single nuclear spins by optical pumping \\
of NV color centers in diamond at room temperature}% Force line breaks with \\

\author{V.~Jacques$^{1,\dag,\ast}$, P.~Neumann$^{1,\dag}$, J.~Beck$^{1}$, M.~Markham$^{2}$, D.~Twitchen$^{2}$, J.~Meijer$^{3}$, F.~Kaiser$^{1}$, G.~Balasubramanian$^{1}$, F.~Jelezko$^{1}$ and J.~Wrachtrup$^{1}$}

\email{v.jacques@physik.uni-stuttgart.de}

\affiliation{$^{1}$3. Physikalisches Institut, Universit$\ddot{\rm a}$t Stuttgart, 70550 Stuttgart, Germany\\
$^{2}$Element Six Ltd., King's Ride Park, Ascot, Berkshire SL5 8BP, UK\\
$^{3}$Central laboratory of ion beam and radionuclides, Ruhr Universit$\ddot{\rm a}$t Bochum, Germany}

\altaffiliation{These authors contributed equally to this work.}

\date{\today}

\begin{abstract}
We report a versatile method to polarize single nuclear spins in diamond, based on optical pumping of a single NV defect and mediated by a level-anti crossing in its excited state. A nuclear spin polarization higher than $98\%$ is achieved at room temperature for the $^{15}$N nuclear spin associated to the NV center, corresponding to $\mu$K effective nuclear spin temperature. We then show simultaneous initialization of two nuclear spins in vicinity to a NV defect. Such robust control of nuclear spin states is a key ingredient for further scaling up of nuclear-spin based quantum registers in diamond.
\end{abstract}

\maketitle

%Note that efficient polarization of isolated nuclear spin ensembles ($10^{4-5}$) have also been observed in single quantum dots by using resonant circularly polarized optical excitation~\cite{Lai_PRL2006,Maletinsky_PRL2007}.\\
 
%\indent Nuclear spins are attractive candidates for solid-state quantum information processing because of their extremely long coherence time resulting from a strong isolation from electronic and vibrational mechanisms that usually lead to decoherence in solid-state systems~\cite{Chuang_Science1997,Kane_Nature1998}. However, their weak interaction with the environment also makes difficult to isolate and manipulate nuclear spins at the single quantum system level. Indeed, conventional nuclear magnetic resonance techniques require the use of macroscopic nuclear spin ensembles to detect measurable signals for quantum computation~\cite{Chuang_Nature2001,Suter_PRL2004,Revue_NMR}.\\
\indent Owing to extremely long coherence time, nuclear spins are attractive candidates for solid-state quantum information processing~\cite{Chuang_Science1997,Kane_Nature1998}. However, the state of individual nuclear spins is difficult to access because of their low magnetic moment~\cite{Revue_NMR}. Recently, room temperature readout of single nuclear spins in diamond has been achieved by coherently mapping nuclear spin states onto the electron spin of a single NV color center~\cite{Jelezko_PRL2004bis,Lukin_Science2006}, which can be optically polarized and read-out with long coherence time~\cite{Fedor_PRL04,Gaebel_NatPhys2006}. This has been the basis for spectacular experiments in quantum information science, including the realization of a nuclear-spin-based quantum register~\cite{Lukin_Science2007} and multipartite entanglement among single spins at room temperature~\cite{Neumann_Science2008}.\\
\indent  However, most of these experiments~\cite{Jelezko_PRL2004bis,Neumann_Science2008} were performed without any deterministic polarization of nuclear spin states. This random initialization unavoidably decreases the success rate of all local operations as $1/2^{N}$ where $N$ is the number of qubits. Deterministic polarization of the nuclear spins thus appears as a crucial step toward development of a scalable diamond based quantum information processing unit.\\
\indent A high degree of polarization of a single nuclear spin in diamond has been demonstrated in Ref.~\cite{Lukin_Science2007}, using a combination of selective microwave excitation and controlled Larmor precession of the nuclear spin state. Another strategy is to use a Level Anti Crossing (LAC) in the ground state~\cite{textbookNMR}, as demonstrated in early ensemble experiments~\cite{Manson_PRB1993}. However, for the present purpose working close to the ground state LAC provides experimental challenges for qubit manipulation resulting from the mixed character of electron and nuclear spin states.\\
\indent In this Letter, we demonstrate a new method to polarize single nuclei in diamond which is simply based on optical pumping. A nuclear spin polarization higher than $98\%$ is achieved, corresponding to $\mu$K effective nuclear spin temperature. A polarization mechanism is given to account for such experimental observation, based on a LAC in the excited state of the NV defect. We then demonstrate simultaneous initialization of two nuclear spins in close vicinity to a NV defect, which provides efficient initialization of a three qubit quantum register by including the electron spin.\\
\indent The NV defect center in diamond consists of a substitutional nitrogen atom (N) associated with a vacancy (V) in an adjacent lattice site (Fig.~1(a)). For the negatively charged NV color center, which is addressed in this study, the ground state is a spin triplet state $^{3} \rm A$, with a zero-field splitting $D_{gs}=2.87$ GHz between spin sublevels $m_{s}=0$ and $m_{s}=\pm 1$~\cite{Manson_PRB2006}. The excited state $^{3} \rm E$ is also a spin triplet, associated with a broadband photoluminescence emission with zero phonon line at $1.945$ eV, which allows optical detection of single NV defects using confocal microscopy. Optical excitation-emission cycles induce a strong  electron spin polarization into the $m_{s}=0$ sublevel. This effect results from spin-selective non-radiative inter-system crossing to a metastable state lying between the ground and excited triplet state~\cite{Manson_PRB2006,Rogers_CondMat2008}. Furthermore, the photoluminescence intensity is higher when the $m_{s}=0$ state is populated, allowing optical detection of spin-rotation of a single NV center by optically detected magnetic resonance (ODMR)~\cite{Gruber_Science1997}. \\
\indent Recently, it has been demonstrated that the excited-state zero-field splitting $|D_{es}|$ is around $1.42$~GHz with parallel principle axis system and an isotropic g-factor identical to the one in the ground state ($g_e\approx 2$)~\cite{Afshalom_QuantPh2008,Neumann_QuantPh2008}. Using such values, an excited state LAC is expected for a magnetic field magnitude on the order of $500$~G~\cite{Afshalom_QuantPh2008,Neumann_QuantPh2008,Epstein_NatPhys2005}. In the following, we use this excited-state LAC to polarize single nuclear spins in diamond.\\
\indent We investigate ultra-pure synthetic type IIa diamond crystals prepared using a chemival vapor deposition process. Single NV color centers were artificially created by implanting $7$ MeV $^{15}\rm N$ ions and by annealing the sample for two hours in vacuum at $800 \ ^{\circ}\rm C$~\cite{Meijer_APL2005,Rabeau_APL2006}. Those NV defects are associated to the $^{15}\rm N$ isotope which is a $I=1/2$ nucleus. The energy splitting resulting from hyperfine interaction between this nuclear spin and the electron spin is $\mathcal{A}_{gs}=-3.05$~MHz in the ground state~\cite{Rabeau_APL2006} (Fig.~1(b)).\\
\begin{figure}[t]
 \centerline{\includegraphics[width=8cm]{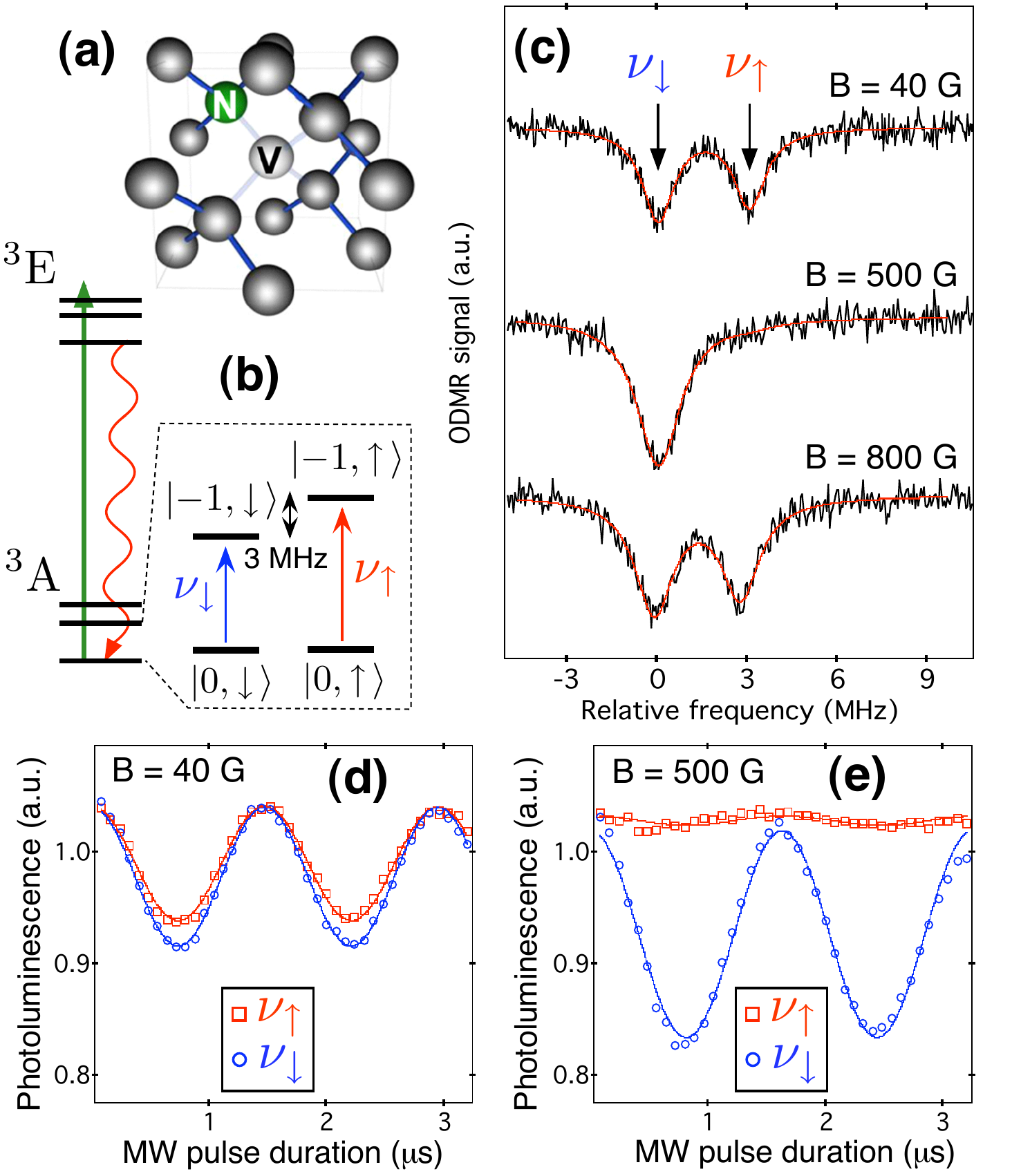}}
    \caption{(color on line). (a)-Atomic structure of the NV defect in diamond. (b)-Simplified energy-level diagram showing the ground state hyperfine structure associated with $^{15}\rm N$ nuclear spin states $\left|\uparrow \right.\rangle$ and $\left|\downarrow \right.\rangle$. Levels associated to electron spin state $m_{s}=+1$ are not shown. Optical transitions (green arrow) are used to polarize and read-out the electron spin state. (c)-ODMR spectra recorded at different magnitudes of a magnetic field applied along the NV symmetry axis ([111] crystal axis). Nuclear spin polarization is observed at the excited-state LAC ($B\approx 500$ G). Solid lines are data fitting using Lorentzian functions. Identical results are obtained for $m_{s}=0$ to $m_{s}=+1$ manifold (data not shown). (d),(e)- Selective Rabi nutation measurements using microwave pulses at frequency $\nu_{\downarrow}$ (blue points) and $\nu_{\uparrow}$ (red points). The experiment is performed for $B=40$ G (d) and $B=500$ G (e). Solid lines are data fitting using cosine functions }
    \end{figure}
\indent ODMR spectra of single NV color centers are recorded by applying microwaves and monitoring the photoluminescence intensity. In addition, a magnetic field ($B$) is applied along the NV symmetry axis ([111] crystal axis). As shown in Fig.~1(c), two electron spin resonances (ESR) are evidenced in ODMR spectra recorded at small magnetic field magnitude, each resonance being associated to a given orientation of the nuclear spin, $\left|\uparrow \right.\rangle$ at frequency $\nu_{\uparrow}$ and $\left|\downarrow \right.\rangle$ at frequency $\nu_{\downarrow}$.  By keeping the magnetic field aligned but increasing its magnitude up to $500$ G, which corresponds to the excited state LAC, the ESR line at frequency $\nu_{\uparrow}$ disappears, indicating a strong polarization of the nuclear spin in state $\left|0,\downarrow \right.\rangle$ (Fig.~1(c))~\cite{NoteHanson}. \\
\indent The nuclear-spin polarization is measured as :
\begin{equation}
\mathcal{P}=\frac{\mathcal{I}(\nu_{\downarrow})-\mathcal{I}(\nu_{\uparrow})}{\mathcal{I}(\nu_{\downarrow})+\mathcal{I}(\nu_{\uparrow})}
\label{DefP}
\end{equation}
where $\mathcal{I}(\nu_{\uparrow})$ (resp. $\mathcal{I}(\nu_{\downarrow})$) is the integral of the ESR peak at frequency $\nu_{\uparrow}$ (resp. $\nu_{\downarrow}$). By fitting each ESR line with Lorentzian functions, we infer a polarization $\mathcal{P}=0.98\pm0.01$. Owing the nuclear-Zeeman splitting between states $\left|0,\uparrow \right.\rangle$ and $\left|0,\downarrow \right.\rangle$ ($\approx 200$~kHz at $B=500$~G), such polarization corresponds to a $\mu$K effective nuclear spin temperature.\\
\indent In order to confirm this observation, selective Rabi nutations are performed for each ESR line using the standard pulse sequence described in Ref.~\cite{Fedor_PRL04}. At low magnetic field magnitude, the contrast of the Rabi nutation is almost identical for each resonance line, demonstrating that the two states $\left|0,\uparrow \right.\rangle$ and $\left|0,\downarrow \right.\rangle$ are populated with similar probabilities. Around the excited-state LAC, the contrast associated to state $\left|0,\uparrow \right.\rangle$ vanishes whereas the one associated to state $\left|0,\downarrow \right.\rangle$ becomes twice higher (Fig.~1(e)). This result constitutes another demonstration of the nuclear spin polarization in state $\left|0,\downarrow \right.\rangle$.\\
\indent We now propose a mechanism to account for the observed nuclear-spin polarization. Assuming that the magnetic field $B$ is perfectly aligned along the NV-axis (z-axis) and neglecting the nuclear Zeeman splitting, the excited-state Hamiltonian is given by 
\begin{equation}
H=D_{es}\hat{S}_z^2 + g_{e}\mu_{B} B \hat{S}_z + \mathcal{A}_{es} \; \hat{\underline{S}} \; \hat{\underline{I}}
\label{Hamilto}
\end{equation}
where $\hat{\underline{S}}$ and $\hat{\underline{I}}$ are the electron and nuclear spin operators, $D_{es}$ the excited-state zero-field splitting, $g_{e}$ the electron g-factor, $\mu_{B}$ the Bohr magneton and $\mathcal{A}_{es}$ the excited-state hyperfine coupling. Recent ESR experiments have shown that hyperfine interaction with a $^{15}$N nuclear spin is $\mathcal{A}_{es}\approx 60$ MHz in the excited-state~\cite{Afshalom_QuantPh2008}. \\ %This is much stronger than in the ground state because the electronic wave function is localized on the nitrogen in the excited-state whereas it is almost entirely on the three carbon dangling orbitals in the ground state~\cite{Twitchen_PRB2008}. Using this Hamiltonian, we also neglect strain-induced splitting of the excited-state spin sublevels. It has been shown recently that such splitting is however much smaller than $D_{es}$ in ultra-pure diamond sample, and as a result does not affect the position of the LAC~\cite{Neumann_QuantPh2008}. \\
\indent We assume $D_{es}$ to be negative~\cite{NotePol} and we restrict the study to the excited-state $m_s~=~0$ and $m_{s}=+1$ sublevels. In the basis $\left[\left|+1,\uparrow \right.\rangle ; \left|+1,\downarrow \right.\rangle ; \left|0,\uparrow \right.\rangle ; \left|0,\downarrow \right.\rangle \right]$ and by choosing the origin of energy at level $\left|0,\uparrow \right.\rangle$, the Hamiltonian described by equation~(\ref{Hamilto}) can be written as 
\[
H=
\left(
\begin{array}{cccc}
	\epsilon_{+1}^{\uparrow}+b&0&0&0\\
	0&\epsilon_{+1}^{\downarrow}+b&a&0\\
	0&a&0&0\\
	0&0&0&0
\end{array}
\right) 
\rm{with}
\ \begin{array}{l}
	\epsilon_{+1}^{\downarrow\uparrow}=D_{es} \pm \mathcal{A}_{es}/2\\
	b=g_e \mu_B B\\
	a=\mathcal{A}_{es}/\sqrt{2} \ .
\end{array}
\]
The eigenstates of such Hamiltonian are $\left|0,\downarrow \right.\rangle$, $\left|+1,\uparrow \right.\rangle$, $\left|+ \right.\rangle=\alpha \left|0,\uparrow \right.\rangle + \beta \left|+1,\downarrow \right.\rangle$ and $\left|- \right.\rangle=\beta \left|0,\uparrow \right.\rangle - \alpha \left|+1,\downarrow \right.\rangle$. The position of the associated eigen energies as well as the values of parameters $\alpha$ and $\beta$ are represented as a function of the magnetic field magnitude in Fig.~\ref{fig:polmechanism}(a). This energy diagram shows an excited-state LAC occuring around $B_{\rm LAC}\approx 500$~G. Note that in the Hamiltonian defined by equation~(\ref{Hamilto}), we have neglected splitting of excited-state spin sublevels caused by local strain on the sample, which might be different from one NV defect to another. It has been shown recently that such strain-induced splitting is however much smaller than $|D_{es}|$ in ultra-pure diamond sample, and as a result does not affect the position of the LAC~\cite{Neumann_QuantPh2008}.\\
\begin{figure}[t]
 \centerline{\includegraphics[width=8cm]{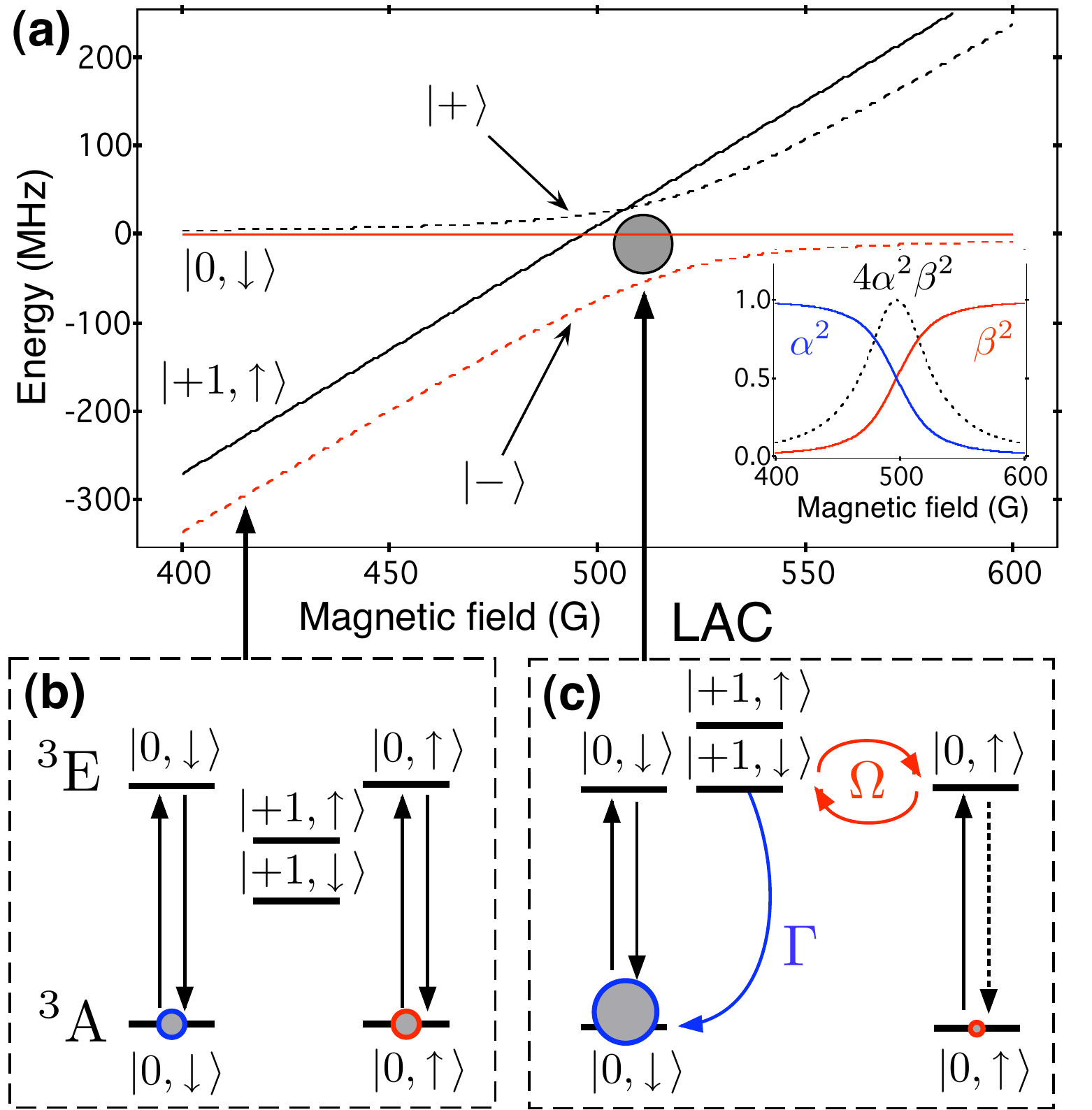}}
    \caption{(color on line). Nuclear spin polarization mechanism. (a)- Eigenstates of the excited-state Hamiltonian described by equation~(\ref{Hamilto}) showing a LAC at $B_{\rm LAC}\approx500$~G. This simulation is performed using $D_{es}=-1420$~MHz and $\mathcal{A}_{es}=+60$~MHz. The inset shows the evolution of parameters $\alpha$, $\beta$ and $4\alpha^{2}\beta^{2}$ as a function of the magnetic field magnitude. (b)- Far from LAC, no nuclear spin polarization is achieved as optical transitions ($^{3}\rm A\rightarrow ^{3}\rm E$) are fully nuclear-spin conserving (black arrows). (c)- At LAC, precession at frequency $\Omega$ between excited-state sublevels $\left|0,\uparrow \right.\rangle$ and $\left|+1,\downarrow \right.\rangle$ can lead to nuclear-spin flip, which can be transferred to the ground state through non-radiative inter-system crossing (blue arrow).}
    \label{fig:polmechanism}
    \end{figure}
\indent At low magnetic field magnitudes, $\alpha\approx 1$ and $\beta\approx 0$. In such regime, optical transitions from the ground to the excited state are fully nuclear spin-conserving as no state mixing in the excited state is occurring (Fig.~\ref{fig:polmechanism}(b)). As a result, the nuclear spin is not polarized.\\
\indent Increasing the magnetic field magnitude close to the LAC, $\alpha$ and $\beta$ begin to balance. The transition from $\left|0,\downarrow \right.\rangle$ to the excited state remains nuclear spin conserving whereas the transition from $\left|0,\uparrow \right.\rangle$ results to $\left(\alpha\left|+ \right.\rangle + \beta\left|- \right.\rangle\right)$ in the excited state. This superposition state then starts to precess between $\alpha\left|+ \right.\rangle + \beta\left|- \right.\rangle=\left|0,\uparrow \right.\rangle$ and $\alpha\left|+ \right.\rangle - \beta\left|- \right.\rangle=(\alpha^2-\beta^2)\left|0,\uparrow \right.\rangle + 2\alpha\beta\left|+1,\downarrow \right.\rangle$ at frequency $
\Omega= 1/2\hbar[(b+\epsilon_{+1}^{\downarrow})^2+4a^2]^{1/2}$, where $\hbar$ is Planck's constant (Fig.~\ref{fig:polmechanism}(c)). The maximum probability $p_{max}(B)$ to find the nuclear spin flipped from $\left|\uparrow \right.\rangle$ to $\left|\downarrow \right.\rangle$ within this precession is given by $p_{max}(B)=4\alpha^2\beta^2$, which follows a Lorentzian magnetic field dependence (see inset of Fig.~\ref{fig:polmechanism}(a)). The nuclear spin-flip has then a probability to be transferred to the ground state sublevel $\left|0,\downarrow \right.\rangle$ by non-radiative inter-system crossing through the metastable singlet state responsible for electron spin polarization of the NV defect~\cite{Manson_PRB2006}. Every subsequent excitation and decay cycles increase spin polarization in state $\left|0,\downarrow \right.\rangle$.\\
    \begin{figure}[b]
 \centerline{\includegraphics[width=8cm]{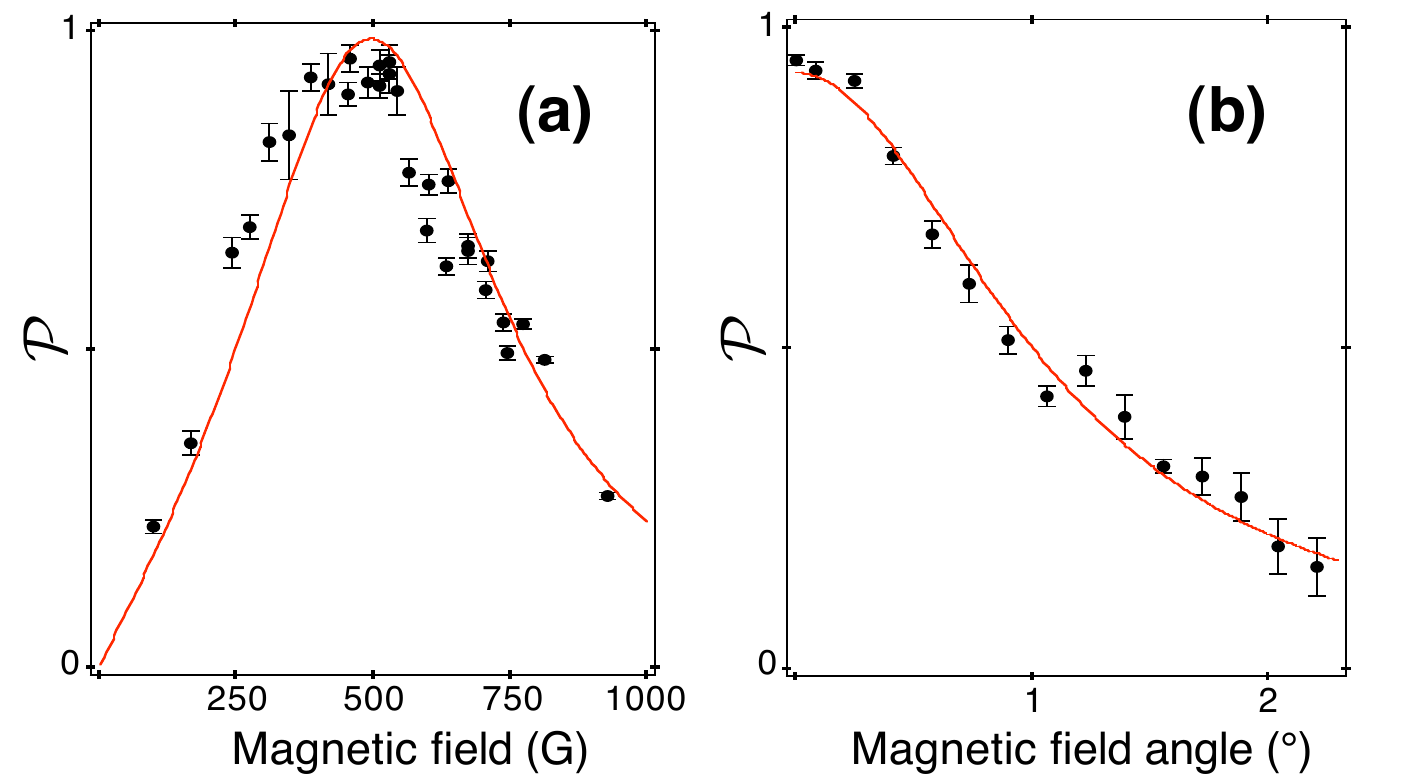}}
    \caption{(color on line). (a)-Nuclear spin polarization $\mathcal{P}$ as a function of the magnetic field magnitude, while keeping its orientation along the NV symmetry axis. $\mathcal{P}$ is estimated by recording ODMR spectra and using Eq.~(\ref{DefP}). Red solid line is data fitting using Eq.~(\ref{Bdepend}) with $k_{eq}^{0}/\Gamma$ as single adjustable parameter ($k_{eq}^{0}/\Gamma=0.009\pm0.001$). (b)-Nuclear spin polarization $\mathcal{P}$ as a function of the magnetic field angle, its magnitude being fixed at $B=472$ G. The solid line is a guide for eyes.}
    	\label{fig:pol}
    \end{figure}
\indent As long as the precession frequency $\Omega$ is on the same order or faster than the excited state decay rate $1/\tau$, the average probability to flip the nuclear spin in the excited-state is $p_+(B)=p_{max}(B)/2$. Using $\tau\approx12$ ns and $\mathcal{A}_{es}\approx 60$~MHz~\cite{Afshalom_QuantPh2008}, this requirement is fullfilled even at the exact position of the LAC where $\Omega$ is minimal. All afore mentioned arguments hold as well for the crossing of the levels $m_s=0$ and $m_s=-1$ when the sign of the magnetic field is reversed, leading to the  probability $p_-(B)=p_+(-B)$ for a spin flip from $\left|\downarrow \right.\rangle$ to $\left|\uparrow \right.\rangle$.\\
\indent Starting with a mixture of $c\left|0,\downarrow \right.\rangle$ and $d\left|0,\uparrow \right.\rangle$ in the ground state, the nuclear spin polarization $\mathcal{P}$ defined by equation~(\ref{DefP}) can be written as $\mathcal{P}=c^{2}-d^{2}$. As rate of polarisation $k_\pm$ and depolarization $k_{eq}$, we assume
\begin{eqnarray}
k_+&=&d^{2}\,p_+\,\Gamma=(1-\mathcal{P})\,p_+\,\Gamma/2\\
k_-&=&-c^{2}\,p_-\,\Gamma=-(1+\mathcal{P})\,p_-\,\Gamma/2\\
k_{eq}&=&-\,k_{eq}^0\,\mathcal{P} \ .
\end{eqnarray}
Here $\Gamma$ is the rate of nuclear-spin conserving intersystem crossing from the excited state to the ground state sublevel $\left|0,\downarrow \right.\rangle$ and $k_{eq}^0$ takes into account all forces that are driving the nuclear spin polarization to its equilibrium state. For such simple model, $\Gamma$ depends on the intensity of the laser which drives the optical transitions. For the steady state, $k_+ + k_- + k_{eq}^{0}=0$, leading to
\begin{equation}
\mathcal{P}(B)=\frac{p_+ - p_-}{\frac{2k_{eq}^0}{\Gamma} + p_+ + p_-} \ .
\label{Bdepend}
\end{equation}
\indent The evolution of polarization $\mathcal{P}$ as a function of the magnetic field magnitude is depicted on Fig.~\ref{fig:pol}(a). The data are well fitted using the function defined by equation~(\ref{Bdepend}) and taking $k_{eq}^{0}/\Gamma$ as single adjustable parameter. The dependence of $\mathcal{P}$ on the magnetic field magnitude is broad, showing that a precise adjustement of the magnetic field magnitude to the LAC is not required, {\it e.g.} $\mathcal{P}\approx95\%$ for $B\approx440$ G. Indeed, even small state mixing in the excited-state leads to efficient nuclear-spin polarization through optical pumping. By saturating the optical transition, the speed of the polarization process is just limited by the parameter $\Gamma$ which is on the order of $4$~MHz according to the metastable state lifetime. We also notice that ground state qubits keep high purity by working at the excited-state LAC, since mixing of ground state sublevels occurs around $B\approx1000$~G~\cite{Epstein_NatPhys2005}. As a result local quantum operations in the ground state would not require to switch off the magnetic field.\\
\indent As depicted in Fig.~\ref{fig:pol}(b), the nuclear spin polarization is however very sensitive to the magnetic field alignment along the NV-axis. Inferring the evolution of the nuclear spin polarization as a function of the magnetic field angle would require more sophisticated numerical simulations of the NV color center spin-dynamics, as all spin-states are partially mixed in that case.
   \begin{figure}[t]
\centerline{\includegraphics[width=8.5cm]{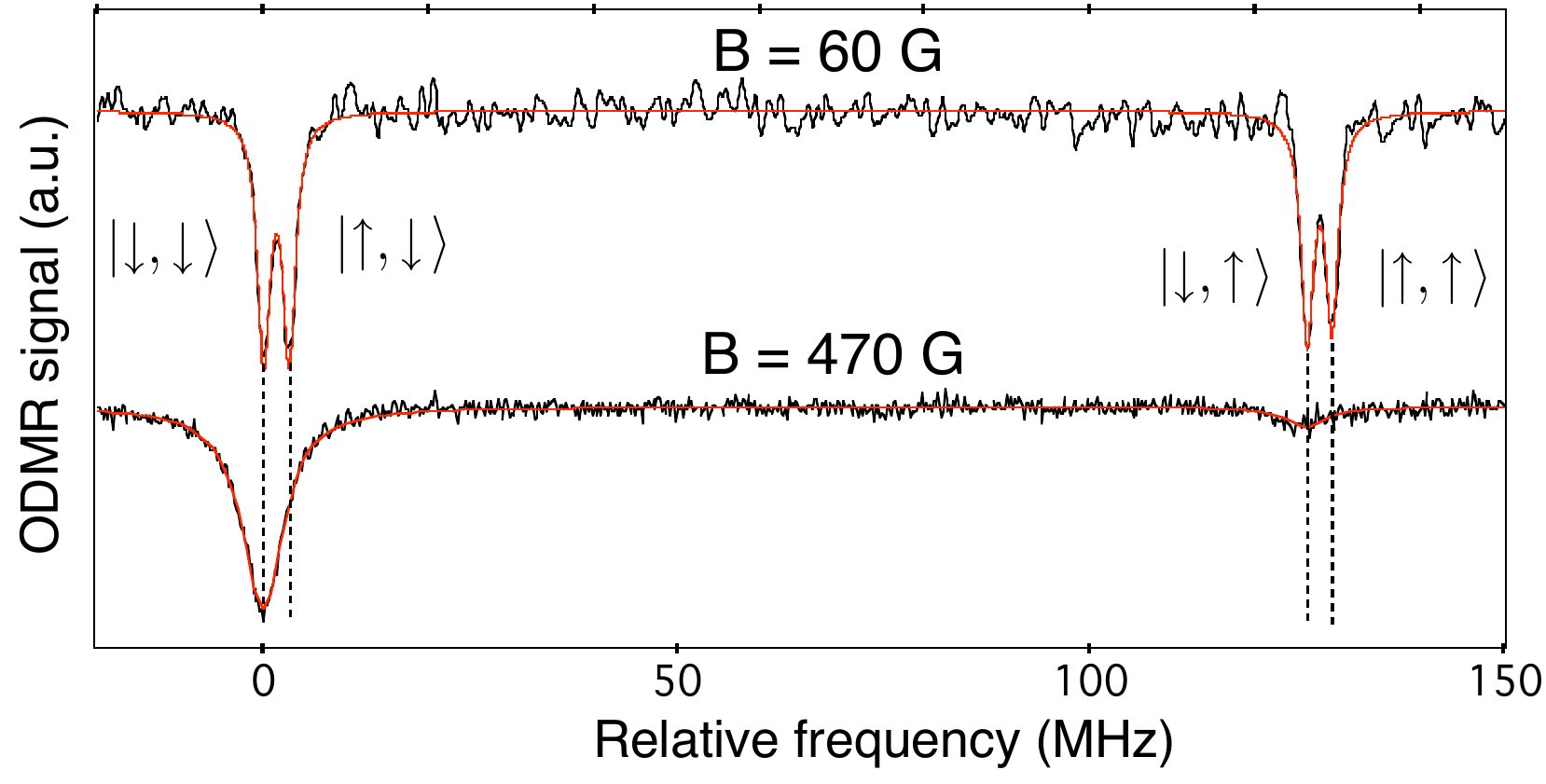}}
    \caption{(color on line). ODMR spectra recorded for a single NV center with a $^{13}\rm C$ on the first coordination shell. At small magnetic field magnitude ($B=60$ G), four ESR lines are observed, each line being associated to a given orientation $\left|^{15}\rm N_{\downarrow \rm or \uparrow},^{13}\rm C_{\downarrow  \rm or\uparrow} \right.\rangle$ of the nuclear spins. Around the LAC ($B=470$ G), $^{15}$N and $^{13}\rm C$ nuclear spins are both polarized. Note that the width of the ESR line is bigger for the measurement performed close to the LAC because of power broadening. However, such width would be small enough to resolve the $3$ MHz hyperfine splitting related to $^{15}$N (see black dot lines). Red solid lines are data fitting using Lorentzian functions.}
   \label{fig:2nuc}
    \end{figure}  
    
\indent As a final experiment, we demonstrate that the nuclear-spin polarization method can be extended to more than one nuclear spin. Fig.~\ref{fig:2nuc} (upper trace) shows the ODMR spectra of a single NV center with a $^{13}\rm C$ on the first coordination shell, leading to a hyperfine splitting around $130$~MHz in the ground state. In such spectrum, four ESR lines are observed, each being associated to a given orientation of $^{13}\rm C$ and $^{15}\rm N$ nuclear spins. By increasing the magnetic field magnitude up to the excited-state LAC, we observe an efficient polarization of both nuclear spins as only one ESR line remains visible (Fig.~\ref{fig:2nuc} lower trace), corresponding to a deterministic initialization of a three qubit quantum register by including the electron spin. From the data we infer a polarization $\mathcal{P}=0.90\pm0.01$ for the double nuclear spin system.\\
\indent Summarizing, we have demonstrated a new method to strongly polarize single nuclear spins in diamond at room temperature, which is experimentally simple to implement since it is only based on optical pumping. Such robust control of nuclear spin states is one of the key ingredients for further scaling up of nuclear-spin based quantum register in diamond~\cite{Neumann_Science2008}.\\
\indent The authors are grateful to R.~Kolesov, L.~Rogers, N.~B.~Manson and R.~Hanson for fruitful discussions. We acknowledge financial support by the European Union (QAP, EQUIND, NEDQIT) and Deutsche Forschungsgemeinschaft (SFB/TR21). V. J. is supported by the Humboldt Foundation.


\begin{thebibliography}{30}
\bibitem{NotePol}
The polarization mechanism also works using $D_{es}<0$ and $D_{gs}<0$. An experimental discrimation between that two possibilities will be examined in future work.
\end{thebibliography}
 \end{document}